\documentclass[12pt,preprint]{aastex}

\slugcomment{Not to appear in Nonlearned J., 45.}
\shorttitle{Strong disk comptonization in GRO J$1655-40$}
\shortauthors{Kubota et al.}

\begin{document}

\title{OBSERVATIONAL EVIDENCE FOR STRONG DISK COMPTONIZATION IN
GRO J$1655-40$}
\author{ Aya {\sc Kubota}\altaffilmark{1} and Kazuo {\sc Makishima}\altaffilmark{2}}\altaffiltext{1}{present address: Institute of Space and Astronautical Science, 3-1-1 Yoshinodai, Sagamihara, Kanagawa 229-8510, Japan}
\altaffiltext{2}{Also Cosmic Radiation Laboratory,
                 Institute of Physical and Chemical Research }
\email{aya@astro.isas.ac.jp}
\affil{Department of Physics,  University of Tokyo, 
7-3-1 Hongo, Bunkyo-ku,\\ Tokyo 113-0033, Japan}

\and

\author{Ken {\sc Ebisawa}\altaffilmark{3}\altaffiltext{3}{Also Universities Space Research Association}}
\affil{code 662, Laboratory of High Energy Astrophysics, NASA/Goddard Space Flight Center, Greenbelt,  MD 20771, U.S.A}

\begin{abstract}
Analysis was made of the multiple {\it RXTE}/PCA data 
on the promised black hole candidate with superluminal jet, 
GRO J$1655-40$, acquired during its 1996--1997 outburst.
The X-ray spectra can be adequately described by the sum of 
an optically thick disk spectrum and a power-law. 
When the estimated 1--100 keV power-law luminosity exceeds 
$1\times10^{37}~{\rm erg~s^{-1}}$ (assuming a distance of 3.2 kpc),
the inner disk radius and  the maximum color temperature 
derived from a simple accretion disk model (a multi-color disk model)
vary significantly with time.
These results reconfirm the previous report by  Sobczak et al. (1999).
In this strong power-law state (once called ``very high state''), 
the disk luminosity decreases with temperature, 
in contradiction to the prediction of the standard Shakura-Sunyaev model. 
In the same state, the intensity of the power-law component correlates 
negatively  with that of the disk component,
and positively  with the  power-law photon index, 
suggesting that the strong power-law is simply the missing 
optically thick disk emission. 
One possible explanation for this behavior
is inverse-Compton scattering in the disk. 
By re-fitting the same data incorporating a disk Comptonization, 
the inner radius and temperature of the underlying disk 
are found to become more constant. 
These results provide one of the first observational confirmations 
of the scenario of disk Comptonization in the strong power-law state.
This strong power-law state seems to appear when color temperature
of the disk exceeds the certain threshold, $\sim$ 1.2 -- 1.3 keV.

\end{abstract}


\section{Introduction}

When the mass accretion rate is high, a black hole binary (BHB) resides in the 
soft state, which is characterized by a very soft spectrum, accompanied by a
power-law tail. 
%
The soft spectral component 
is believed to be thermal emission 
from an optically thick accretion disk around a central BH
(e.g., Makishima et al. 1986),
because it can be reproduced by
a multi-color disk model (MCD model, Mitsuda et al. 1984)
which approximates emission from
a standard accretion disk (Shakura \& Sunyaev 1973).
The model has two parameters; 
the maximum color temperature of the disk, $T_{\rm in}$, and 
an apparent inner radius $r_{\rm in}$. 
Using a spectral hardening factor of $\kappa\simeq 1.7$--2.0 
(Shimura \& Takahara 1997), and 
a correction factor for the inner boundary condition, 
$\xi=0.41$ (Kubota et al. 1998), 
$r_{\rm in}$ can be related to the true inner radius $R_{\rm in}$, as 
\begin{equation}
R_{\rm in}=\kappa ^2 \cdot \xi \cdot r_{\rm in}~~~.
\label{eq:xi}
\end{equation}
X-ray observations indicate that $R_{\rm in}$ remains constant at 
$6R_{\rm g}$ that is the last stable orbit for a non-spinning BH
($R_{\rm g}$ is the gravitational radius).

Although this ``standard picture''
remained generally successful
(e.g., Ebisawa et al. 1993, Tanaka \& Lewin 1995),
recently two deviations from its predictions have been
noticed in some soft-state BHBs.
One is quite small values of $R_{\rm in}$, compared to $6R_{\rm g}$,
and the other is 
significant variations in $R_{\rm in}$.
The former 
is found in the superluminal jet sources,
GRO J$1655-40$ and GRS $1915+105$ (e.g., Zhang et al. 1997).
The latter includes GRO J$1655-40$ (Sobczak et al. 1999; hereafter S99),
LMC X-1 (Wilms et al. 2001) 
and XTE J$1550-564$ (Kubota 2001).
Although these anomalies are often attributed, e.g., to strong
disk Comptonization and/or very high value of $\kappa$,
no convincing evidence has been available.
These issues may be related 
to the general theoretical belief that 
the standard-disk picture is valid only for a rather
limited range of the mass accretion rate (e.g., Esin et al. 1997).

In order to examine to what extent the standard picture is valid,
we have analyzed X-ray spectra of GRO J$1655-40$ 
obtained by multiple pointings with {\it RXTE}. 
This BHB has been reported to exhibit the two peculiarities mentioned above.
Moreover, its BH mass, distance, and inclination angle 
are accurately estimated to be
$7.02\pm0.22~M_\odot$, $3.2\pm0.2$ kpc, and $69.5\pm0.08^\circ$,
respectively (e.g., Orosz \& Bailyn 1997). 
These make GRO J$1655-40$ ideal for our purpose.

\section{Observation and data reduction}
We analyzed 72 archival data sets of GRO J$1655-40$ obtained with the
{\it RXTE}/PCA, 
covering the 16-months outburst in 1996--1997. 
This is the same dataset as utilized by S99
except for the last
few observations when the source exhibited a signature of the hard state.
We co-added the data from the individual proportional counter units,
and produced one co-added spectrum for each pointing.
We select good data in the basic manner for bright sources.

We estimated the PCA background for each observation using 
{\it pcabackest} version 2.1e.
In order to correct a possible $<10$ \% over/under-estimation
of the background by the {\it pcabackest},
we compared the on-source spectra and the predicted model background spectra
in the hardest energy band (60--100 keV),
where the signal flux is usually negligible.
When necessary, we adjusted 
the normalization factor of the background spectrum by $-10\sim+10\%$. 
We make the response matrix for each observation using 
{\it pcarsp} version 7.10, and 
add 1\% systematic error to each energy bin of the PCA spectrum. 
Over the 20--35 keV range, we increased the systematic error to 10\%, 
to take into account the response uncertainties
associated with the Xe-K edge at $\sim30$ keV.

\section{Data analysis and results}
\subsection{Standard modeling}
According to the canonical spectral modeling,
we fit the obtained 3--30 keV PCA spectra of GRO J$1655-40$
with the MCD plus a power-law.
After Yamaoka et al. (2001), we subject the MCD component to several
absorption features (a line and edges).
We do not discuss these absorption features any further.
To the two constituent continuum components, 
we apply a common photoelectric absorption with the
column fixed at $N_{\rm H}=7\times10^{21}~{\rm cm^{-2}}$,
by referring to the {\it ASCA}/GIS data of
this source on 1997 Feb.26 (Kubota 2001).
Moreover, as for the data obtained before 1996 May 21 (MJD 50224), 
we fix the power-law photon index $\Gamma$ 2.1, because 
the power-law component was too weak to constrain $\Gamma$. 

The fits have been acceptable for all the PCA spectra.
In Fig.1, we show evolution of the best-fit model parameters,  
including the disk bolometric luminosity,
$L_{\rm disk}(\equiv 4\pi \sigma r_{\rm in}^2 T_{\rm in}^4$),
the 1--100 keV power-law luminosity, $L_{\rm pow}$, calculated assuming an
isotropic emission,
and their sum, $L_{\rm tot}$. 
Thus, the entire PCA data span of this source can be
divided into three characteristic periods, referring mainly to $L_{\rm pow}$. 
The 1st period (Period 1; before day 141) is characterized by 
a very low level of $L_{\rm pow}$,
while it is very high ($> 1\times 10^{37}~{\rm erg~s^{-1}}$)
in Period 2, 
which was once called ``very high state'' by S99.
In Period 3 (after the data gap), $L_{\rm pow}$ returns low. 

In Period 3, the values of $R_{\rm in}$, 
which are obtained by utilizing eq.(1) with $\kappa=1.7$ and $\xi=0.41$, 
remain constant (26 km)
against relatively large intensity variation,
while in Period 2 they are observed to change
significantly between $\sim6$ km and $\sim24$ km.
When we fix $R_{\rm in}$ at 26 km 
and instead allow 
$N_{\rm H}$ to vary,
the fits for the Period 1 and 2 data become significantly worse
(Fig.1e),
and $N_{\rm H}$ changes violently. 
Thus, the variation of $R_{\rm in}$ is real
as long as we utilize the canonical 
two components model
with constant values of $\kappa$ and $\xi$.
Clearly, GRO~J$1655-40$ in Period 2 violates the standard picture.

In order to highlight the anomalies of GRO J$1655-40$,
we plot $L_{\rm disk}$ against $T_{\rm in}$ in Fig.2a. 
For comparison, we also plot the data points of a typical BHB, LMC X-3 
(e.g., Kuiper et al. 1988),
of which the BH mass (5--7.2~$M_\odot$)
and the inclination angle (65$^\circ$--69$^\circ$) 
are both quite similar to those of GRO J$1655-40$.
The results on LMC X-3 were obtained
through the spectral fitting of the same PCA data as
reported by Wilms et al. (2001).
Thus, the data points for LMC X-3 follow a simple relation of
$L_{\rm disk}\propto T_{\rm in}^4$ with a constant $R_{\rm in}$
as $L_{\rm disk}$ varied by a factor of 10.
Moreover, 
assuming its distance and inclination angle as 55 kpc 
and $66^\circ$ respectively, 
the absolute value of $R_{\rm in}$ 
is calculated as $\sim60$ km, which coincide with $6R_{\rm g}$ 
for a BH of $6M_\odot$. 
In other words, the accretion disk in LMC X-3 
perfectly satisfies the standard picture.
In contrast, GRO J$1655-40$ exhibits a distinct behavior on this
$T_{\rm in}$-$L_{\rm disk}$ plane.
The data points in Period 2 and Period 1 
deviate from the standard $L_{\rm disk}\propto T_{\rm in}^4$ relation,
while those of Period 3 satisfy the relation
except that the value of $R_{\rm in}$ is much smaller than that of LMC X-3.

Although the deviation from the standard picture
has been found in both Period 1 and 2, 
it is much more significant in Period 2 than in Period 1.
In addition, the observed PCA spectra in Period 1 are relatively similar to
those in Period 3 (see S99), while those in Period 2 are characterized by
very strong hard emission component. 
Therefore in this letter, 
we mainly focussed on the anomalous behavior in Period 2. 
Hereafter, we call Period 2 {\it anomalous regime},
while call Period 3 {\it standard regime}. 

\subsection{Differences between {\it anomalous} and {\it standard} regimes}
A prominent difference between the
{\it anomalous} and {\it standard} regimes is found in behavior of the 
power-law component. 
In the {\it anomalous regime}, 
$L_{\rm pow}$ 
negatively correlates to $L_{\rm disk}$, 
in such a way that $L_{\rm tot}$ 
is kept approximately constant at
a ceiling value of $\sim1.7\times10^{38}~{\rm erg~s^{-1}}$ (Fig.1a). 
To our surprise, this ceiling corresponds roughly to $\sim15$ \% of 
the Eddington luminosity, 
$L_{\rm E}\sim 1.1\times10^{39}~{\rm erg~s^{-1}}$
(assuming solar abundance)
for a $\sim7 M_\odot$ BH in GRO J$1655-40$, instead of $L_{\rm E}$ itself.
Furthermore, as shown in Fig.3, $\Gamma$ 
gradually increases 
as $L_{\rm pow}$ gets higher in the
{\it anomalous regime}, while it stays constant at $\sim$2.1 in the
{\it standard regime}.
Therefore, the property of the hard component in the
{\it anomalous regime} may be intrinsically different from
that in the {\it standard regime}.

A simple interpretation of 
the source behavior in the {\it anomalous regime} 
is to presume
that there emerges a third spectral component, 
which is harder than the MCD emission
but softer than the hard component in the {\it standard regime}.
Then, the strong anti-correlation between $L_{\rm pow}$ and $L_{\rm disk}$,
seen in the {\it anomalous regime}, 
can be explained by assuming that this
third component strongly
and negatively correlates with the MCD component.
It is therefore natural to assume that some fraction of the photons 
emitted from the optically thick accretion disk 
are converted into
the third spectral component, instead of directly reaching us.
The third component is most probably 
produced through inverse-Compton scattering of the MCD photons
by high energy electrons that may reside around the disk.

\subsection{Spectral fitting incorporating a Comptonized component}

We re-fit the same PCA spectra in the {\it anomalous regime},
with a three-component model, obtained by adding a Comptonized blackbody
(``{\it compbb}''; Nishimura, Mitsuda, \& Itoh 1986)
component to the original two component model.
The {\it compbb} model has four parameters;
blackbody temperature $T_{\rm bb}$, electron temperature $T_{\rm e}$,
Compton optical depth $\tau$,  and radiative area 
of the blackbody for an isotropic emission.
However, we cannot constrain all these additional parameters,
since the previous two-component model has given acceptable fits.
We accordingly tie $T_{\rm bb}$ to $T_{\rm in}$, assuming the seed photons
for the inverse-Compton to be supplied by the optically thick accretion disk.
We fix $\Gamma$ of the original power-law component to 2.1,
an average 
in the {\it standard regime} (Fig.3).
Furthermore, to avoid a strong coupling between $\tau$ and $T_{\rm e}$,
we fix $T_{\rm e}$ at a representative value of 10 keV,
considering that
strong Compton cooling by ample seed soft photons from the
optically-thick disk will make $T_{\rm e}$ significantly lower than
in the hard state ($T_{\rm e}\sim$30--50 keV; e.g., Grove et al. 1998).

In Fig.1c, we plot the re-estimated $T_{\rm in}$ with
open circles.
By considering the {\it compbb} component, 
the highly deviated data points
in terms of $T_{\rm in}$ have thus settled
back to a smooth long-term trend.
We also re-estimate the luminosity in Fig.2b as
$L_{\rm disk}+L_{\rm cbb}$, where $L_{\rm cbb}$ is the
estimated 0.1--100 keV {\it compbb} luminosity, assuming 
an isotropic emission. 
Thus, $L_{\rm disk}+L_{\rm cbb}$ plotted against the revised $T_{\rm in}$
approximately recovers the standard $L_{\rm disk}\propto {T_{\rm in}}^4$ 
relation for optically-thick accretion disks.
Consequently, the value of $R_{\rm in}$ can also be considered
to remain relatively stable,
even when a significant fraction of the MCD
photons is Comptonized.
We conclude that some part of the power-law seen in
the {\it anomalous regime} 
has the origin in the MCD photons, modified through 
the inverse-Compton process, 
and that the violent variations in the MCD parameters, 
on time scales of few days or shorter, is not real but apparent.

Figure 4a shows the typical PCA spectrum of GRO J$1655-40$ in the 
{\it anomalous regime}, which 
corresponds to Observation A (1996 Aug. 6; day 218) presented in Fig.1a, 
fitted with the three-component model.
We also show the result from the previous two-component fit in Fig.4b, 
where the best-fit model is obtained in the range of 3--30 keV.
although the spectrum is shown in 3--50 keV.


\section{Discussion}

In \S3, we have shown that the scenario of the strong disk Comptonization 
successfully explains the {\it anomalous regime} of GRO J$1655-40$, 
and that the underlying disk really satisfies the 
standard picture. 
A very similar phenomenon has been observed in another Galactic jet source, 
XTE J$1550-564$ (Kubota 2001).
Although the disk Comptonization has been discussed extensively 
in the literature, our results provide one of 
the first unambiguous observational confirmations of such a picture. 
Then, what causes such a strong Comptonization?
As mentioned in \S1,
it was theoretically pointed out 
that the standard disk cannot be stable under high accretion rates.
The {\it anomalous regime} can be hence considered to
occur when the accretion rate reaches a certain upper critical level, 
as indicated by the location of {\it anomalous regime} (upper right) in 
Fig.2b.

%
%

We however remember that the critical disk luminosity, at which
the {\it anomalous} behavior of GRO J$1655-40$ appears, is
only  $\sim15$\% of $L_{\rm E}$,
and that LMC X-3 exhibits no such deviation
from the standard-disk even with $L_{\rm disk}$ close to $L_{\rm E}$.
Another important difference between these two BHBs is that 
the value of $R_{\rm in}$ for GRO J$1655-40$ 
is only $\sim 2R_{\rm g}$ for a $7M_\odot$ BH 
even in the {\it standard regime} , 
while that of LMC X-3 agrees with $6R_{\rm g}$.  
This fact makes the
values of $T_{\rm in}$ for GRO J$1655-40$ much higher than those 
for LMC X-3 when compared at the similar luminosity.
Interestingly, the observed value of the temperature of 
GRO J$1655-40$, $\sim1.2$ keV, 
which corresponds to the critical luminosity, is almost the same as the
upper limit of $T_{\rm in}(\sim1.3$ keV) for LMC X-3. 
We hence suggest that the endpoint of the {\it standard regime} is not
determined by the luminosity but instead by the
temperature, which is thought to be $\sim 1.2$--1.3 keV for both
LMC X-3 and GRO J$1655-40$. 
Such a consideration is consistent with a theoretical expectation that, 
at a high temperature, 
the opacity of the disk is given by electron scattering instead of 
photo-electric absorption that dominates at lower temperatures.

Thus, we consider that the systematically higher $T_{\rm in}$ (or smaller 
$R_{\rm in}$) 
is responsible for 
the anomalous behavior of GRO J$1655-40$. 
Because the innermost stable orbit
becomes smaller for a prograde rotation around a spinning BH,
down to $\sim R_{\rm g}$ in the extreme case, 
the anomalous behavior of GRO J$1655-40$ may be attributed 
ultimately to its BH spin, as already suggested by Zhang et al. (1997), 
and Makishima et al. (2000).


We would like to thank Prof. H. Inoue and Prof. S. Mineshige
for helpful discussions. 
We are also grateful to Dr. C. Done for valuable comments and 
discussions.

\begin{figure}[htbp]
\begin{center}
\includegraphics[clip=true]
{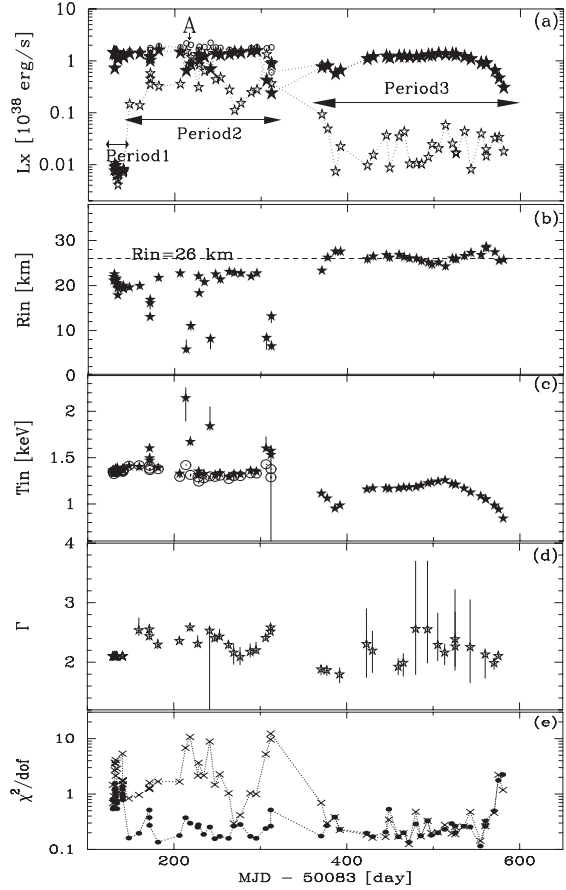}
\caption{
Long-term variation of the spectral parameters of GRO J$1655-40$.
(a) Histories of $L_{\rm disk}$ (filled star),
$L_{\rm pow}$(open star),
and $L_{\rm tot}$ (open circle), 
The three characteristic periods are indicated in the top panel.
(b)--(d) Time histories of $r_{\rm in}$, $T_{\rm in}$, and $\Gamma$,
respectively. 
Large circles in panel (c) represent $T_{\rm in}$ obtained by incorporating 
the disk-Comptonization. (e) That of $\chi^2/{\rm dof}$;
filled circles represent $\chi^2/{\rm dof}$ with 
the free-$R_{\rm in}$ and fixed-$N_{\rm H}$ fits,
whereas crosses those when $R_{\rm in}$ is fixed at 26 km and 
$N_{\rm H}$ is left free.}
\label{fig1}
\end{center}
\end{figure}

\begin{figure}[htbp]
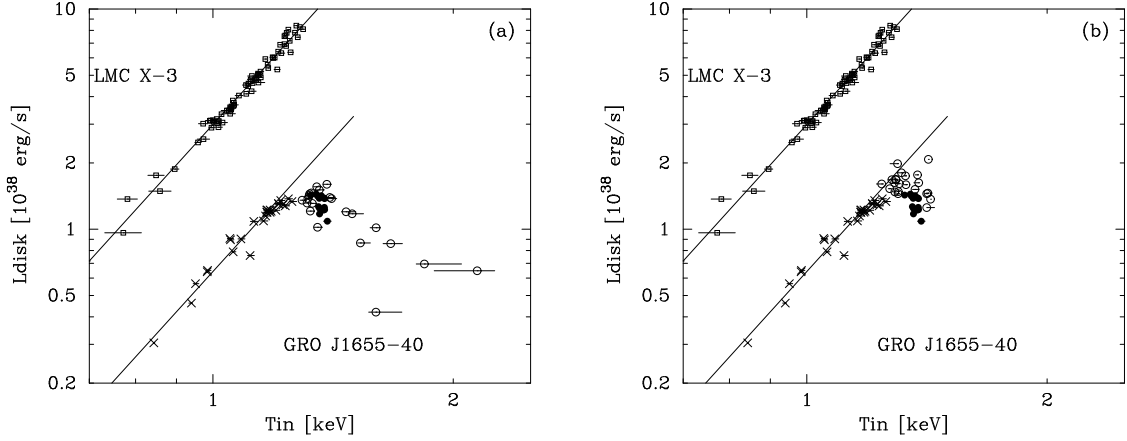

\begin{center}
\begin{minipage}{0.47\textwidth}
\includegraphics[clip=true,height=0.9\textwidth,angle=-90]
{./fig2a.ps}
\end{minipage}
\begin{minipage}{0.47\textwidth}
\includegraphics[clip=true,height=0.9\textwidth,angle=-90]
{./fig2b.ps}
\end{minipage}
\caption{(a) The calculated $L_{\rm disk}$ plotted against the observed
$T_{\rm in}$. As for GRO J$1655-40$, three kinds of symbols specify the data
obtained during Period 1 (filled circles), Period 2 (open circles),
and Period 3 (crosses). The results of LMC X-3 (open squares) are also
plotted for comparison.
The solid lines represent the $L_{\rm disk}\propto T_{\rm in}^4$ relation. 
(b) Same as in panel (a), but the data points in Period 2
({\it anomalous regime}) of GRO J$1655-40$ are 
re-calculated considering the Comptonized component as 
$L_{\rm disk}+L_{\rm cbb}$.}
\label{fig2}
\end{center}
\end{figure}

\begin{figure}[htbp]
\begin{center}
\includegraphics[clip=true, angle=-90,width=6cm]
{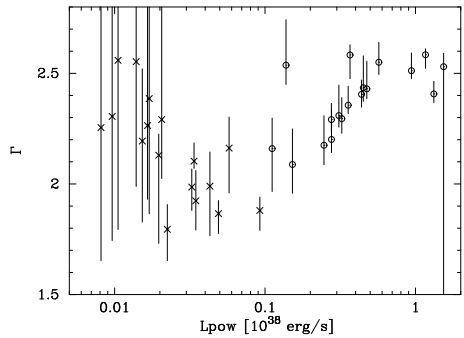}
\caption{The observed $\Gamma$ plotted against $L_{\rm pow}$.
Symbols are the same as in Fig.2.}
\label{fig3}
\end{center}
\end{figure}

\begin{figure}[htbp]
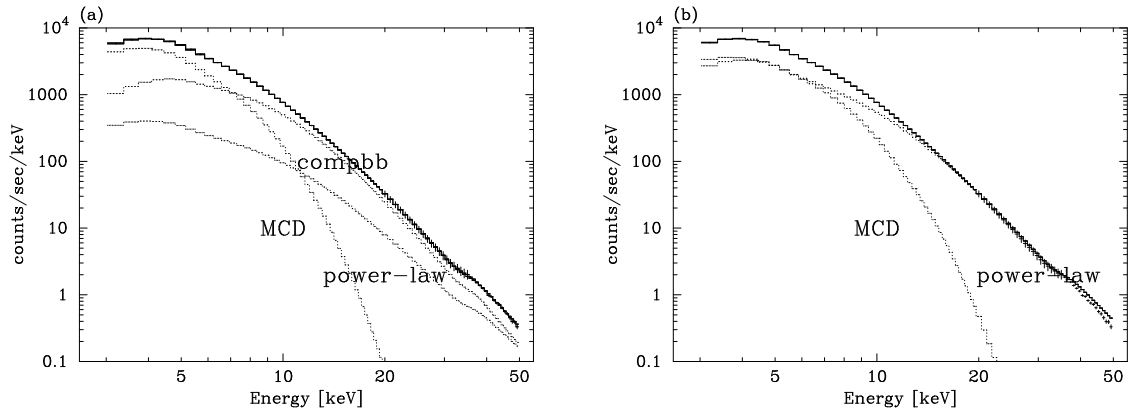

\begin{center}
\begin{minipage}{0.47\textwidth}
\includegraphics[clip=true,height=0.9\textwidth,angle=-90]
{./cbb.ps}
\end{minipage}
\begin{minipage}{0.47\textwidth}
\includegraphics[clip=true,height=0.9\textwidth,angle=-90]
{./mcd.ps}
\end{minipage}
\caption{The PCA spectrum of GRO J$1655-40$ obtained in Observation A in 
Fig.1a. 
Predictions of the best-fit three-component model (panel a) and 
the canonical two-component model (panel b) are also shown, 
together with individual components.
}
\label{fig4}
\end{center}
\end{figure}


\end{document}